\input harvmac
\input epsf
\input amssym


\def\CN{{\cal N}}

\def\CS{{\cal S}}
\def\hat{\widehat}

\def\lfm#1{\medskip\noindent\item{#1}}
\def\Tr{{\rm Tr}}

\def\sqr#1#2{{\vcenter{\vbox{\hrule height.#2pt
    \hbox{\vrule width.#2pt height#1pt \kern#1pt
    \vrule width.#2pt}
    \hrule height.#2pt}}}}
\def\square{\mathchoice\sqr65\sqr65\sqr{2.1}3\sqr{1.5}3}

\lref\MinahanFG{
  J.~A.~Minahan, D.~Nemeschansky,
  ``An N=2 superconformal fixed point with E(6) global symmetry,''
Nucl.\ Phys.\  {\bf B482}, 142-152 (1996).
[hep-th/9608047].
}

\lref\MinahanCJ{
  J.~A.~Minahan, D.~Nemeschansky,
  ``Superconformal fixed points with E(n) global symmetry,''
Nucl.\ Phys.\  {\bf B489}, 24-46 (1997).
[hep-th/9610076].
}

\lref\GanorXD{
  O.~J.~Ganor,
  ``Toroidal compactification of heterotic 6-d noncritical strings down to
  four-dimensions,''
  Nucl.\ Phys.\  B {\bf 488}, 223 (1997)
  [arXiv:hep-th/9608109].
}

\lref\GukovKT{
  S.~Gukov and A.~Kapustin,
  ``New N=2 superconformal field theories from M / F theory orbifolds,''
  Nucl.\ Phys.\  B {\bf 545}, 283 (1999)
  [arXiv:hep-th/9808175].
}

\lref\GaiottoWE{
  D.~Gaiotto,
  ``N=2 dualities,''
[arXiv:0904.2715 [hep-th]].
}

\lref\BeniniMZ{
  F.~Benini, Y.~Tachikawa, B.~Wecht,
  ``Sicilian gauge theories and N=1 dualities,''
JHEP {\bf 1001}, 088 (2010).
[arXiv:0909.1327 [hep-th]].
}

\lref\GaiottoGZ{
  D.~Gaiotto, J.~Maldacena,
  ``The Gravity duals of N=2 superconformal field theories,''
[arXiv:0904.4466 [hep-th]].
}

\lref\TachikawaTT{
  Y.~Tachikawa, B.~Wecht,
  ``Explanation of the Central Charge Ratio 27/32 in Four-Dimensional Renormalization Group Flows between Superconformal Theories,''
Phys.\ Rev.\ Lett.\  {\bf 103}, 061601 (2009).
[arXiv:0906.0965 [hep-th]].
}

\lref\IW{
K.~Intriligator and B.~Wecht,
``The exact superconformal R-symmetry maximizes a,''
Nucl.\ Phys.\ B {\bf 667}, 183 (2003)
[arXiv:hep-th/0304128].
}

\lref\HeckmanQV{
  J.~J.~Heckman, Y.~Tachikawa, C.~Vafa, B.~Wecht,
  ``N = 1 SCFTs from Brane Monodromy,''
JHEP {\bf 1011}, 132 (2010).
[arXiv:1009.0017 [hep-th]].
}

\lref\HeckmanHU{
  J.~J.~Heckman, C.~Vafa, B.~Wecht,
  ``The Conformal Sector of F-theory GUTs,''
JHEP {\bf 1107}, 075 (2011).
[arXiv:1103.3287 [hep-th]].
}

\lref\MaldacenaMW{
  J.~M.~Maldacena, C.~Nunez,
  ``Supergravity description of field theories on curved manifolds and a no go theorem,''
Int.\ J.\ Mod.\ Phys.\  {\bf A16}, 822-855 (2001).
[hep-th/0007018].
}

\lref\SeibergBZ{
  N.~Seiberg,
  ``Exact results on the space of vacua of four-dimensional SUSY gauge theories,''
Phys.\ Rev.\  {\bf D49}, 6857-6863 (1994).
[hep-th/9402044].
}

\lref\KomargodskiVJ{
  Z.~Komargodski, A.~Schwimmer,
  ``On Renormalization Group Flows in Four Dimensions,''
[arXiv:1107.3987 [hep-th]].
}

\lref\TachikawaEA{
  Y.~Tachikawa and K.~Yonekura,
  ``N=1 curves for trifundamentals,''
  JHEP {\bf 1107}, 025 (2011)
  [arXiv:1105.3215 [hep-th]].
}

\lref\GreenDA{
  D.~Green, Z.~Komargodski, N.~Seiberg, Y.~Tachikawa, B.~Wecht,
  ``Exactly Marginal Deformations and Global Symmetries,''
JHEP {\bf 1006}, 106 (2010).
[arXiv:1005.3546 [hep-th]].
}

\lref\IbouTA{
I.~Bah, {\it to appear}
}

\lref\GauntlettZH{
  J.~P.~Gauntlett, D.~Martelli, J.~Sparks and D.~Waldram,
  Class.\ Quant.\ Grav.\  {\bf 21}, 4335 (2004)
  [arXiv:hep-th/0402153].
}

\lref\CsakiUJ{
  C.~Csaki, P.~Meade, J.~Terning,
  ``A Mixed phase of SUSY gauge theories from a maximization,''
JHEP {\bf 0404}, 040 (2004).
[hep-th/0403062].
}

\lref\GaiottoJF{
  D.~Gaiotto, N.~Seiberg, Y.~Tachikawa,
  ``Comments on scaling limits of 4d N=2 theories,''
JHEP {\bf 1101}, 078 (2011).
[arXiv:1011.4568 [hep-th]].
}

\lref\CraigTX{
  N.~Craig, R.~Essig, A.~Hook, G.~Torroba,
  ``New dynamics and dualities in supersymmetric chiral gauge theories,''
JHEP {\bf 1109}, 046 (2011).
[arXiv:1106.5051 [hep-th]].
}

\lref\CraigWJ{
  N.~Craig, R.~Essig, A.~Hook, G.~Torroba,
  ``Phases of N=1 supersymmetric chiral gauge theories,''
[arXiv:1110.5905 [hep-th]].
}

\

\def\drawbox#1#2{\hrule height#2pt
             \hbox{\vrule width#2pt height#1pt \kern#1pt \vrule 
width#2pt}
                   \hrule height#2pt}

\def\Asym#1#2{\vcenter{\vbox{\drawbox{#1}{#2}
                   \kern-#2pt       
                   \drawbox{#1}{#2}}}}

\Title{\vbox{\baselineskip12pt
\hbox{MCTP-11-39} }}
{\vbox{\centerline{New $\CN=1$ Superconformal Field Theories} \medskip \centerline{In Four Dimensions}
}}
\centerline{ Ibrahima Bah$^1$ and Brian Wecht$^{1,2}$}
\bigskip
\centerline{${}^1\,$Michigan Center for Theoretical Physics}
\centerline{University of Michigan}
\centerline{Ann Arbor, MI 48109 USA}
\bigskip
\centerline{${}^2\,$Center for the Fundamental Laws of Nature}
\centerline{Harvard University}
\centerline{Cambridge, MA 02138 USA}

\bigskip
\noindent
We construct a large new family of four-dimensional $\CN=1$ superconformal field theories by coupling Gaiotto's $T_N$ theories to $\CN=1$ vector multiplets. In particular, we consider theories in which various $T_N$ blocks are linked together via bifundamental and adjoint chiral superfields, with no superpotential. We find that while some of these constructions appear to give new strongly coupled SCFTs, others lead to violations of the $a$-theorem, and thus do not appear to be good interacting theories.

\Date{November 2011}

\newsec{Introduction}

The infrared phases of four-dimensional quantum field theories are notoriously difficult to study. Even if a theory starts off with a weakly coupled description in the ultraviolet, it can (and often does) then flow to some strongly coupled phase at long distances. The particular details of the infrared phase are often not at all obvious from the free UV description, and even when we know the answer (as in QCD), the underlying dynamics can still be wildly out of theoretical control. We thus often think of the UV Lagrangian as something that is given to us, and relegate our ignorance to the mysteries of the infrared.

It has been known for some time that it is occasionally possible to make detailed statements about quantum field theories even when a weakly coupled UV Lagrangian is not available. When the theory is in a conformal phase, the constraints provided by the additional symmetries allow for exact results which are often not accessible in non-conformal theories. This increased power makes a Lagrangian useful but not strictly necessary, depending on the questions in which one is interested. For example, there are well-known such cases in two dimensions (e.g.  minimal models) where we can say many things without any reference to a Lagrangian. 

Four-dimensional CFTs are not nearly as constrained as those in two dimensions, since the conformal algebra no longer has infinite dimension. However, there are still many theories in which we can make progress without a weakly coupled description, as long as we have enough supersymmetry. The most famous such cases are probably the $\CN=2$ SCFTs found by Minahan and Nemeschansky in \refs{\MinahanFG,\MinahanCJ}. These theories have $E_n$ global symmetries, and their Seiberg-Witten curves are known, along with the exact scaling dimensions of various operators. Additionally, these theories are strongly coupled (as one can read off from the SW curve), and have no known weakly coupled UV descriptions.
Some related examples were found in \refs{\GanorXD,\GukovKT}. 

The number of ``non-Lagrangian" CFTs saw a dramatic boost with Gaiotto's discovery of $T_N$ theories in \GaiottoWE. These $\CN=2$ SCFTs come from M5-branes wrapping a thrice-punctured Riemann surface. The resulting theories have an $SU(N)^3$ global symmetry and do not have any known weakly coupled UV avatars. Despite this lack of information, we do (as in the $E_n$ SCFTs) know a surprising amount about these theories, including many of the gauge-invariant operators in the theory, as well as their scaling dimensions and charges under  global symmetries. 

One reason we know so much about $E_n$ and $T_N$ theories is because these theories have $\CN=2$ SUSY. Naively, theories with less supersymmetry might seem inaccessible to study. This is, however, not the case. If we start with a theory with $\CN=2$ SUSY and then break to $\CN=1$, we can often still make exact statements. In particular, if we want to know the scaling dimensions of  chiral primary operators, the only thing we need access to is information about global symmetries: As long as we break SUSY in such a way that we can track which global symmetries are unbroken, we can use $a$-maximization \IW\ to compute the $\CN=1$ superconformal R-symmetry.\foot{This is the philosophy adopted in \refs{\HeckmanQV, \HeckmanHU}, where such computations were performed for $\CN=1$ deformations of $\CN=2$ $E_n$ SCFTs.} The dimension $\Delta$ of a chiral primary ${\cal O}$ is then given by $\Delta({\cal O}) = {3 \over 2} R({\cal O})$.

$\CN=1$ deformations of $T_N$ theories were first studied in \BeniniMZ. In that work, the authors gauge the $SU(N)^3$ global symmetry, and then build theories by connecting $T_N$'s through these gauge groups. SUSY is then broken by adding a mass term for the $\CN=1$ adjoint chiral superfields in the $\CN=2$ vector multiplets. Integrating out the adjoints results in a superpotential which uniquely fixes the R-symmetry, so $a$-maximization is unnecessary. The original $\CN=2$ theories were shown in \GaiottoGZ\ to be dual to the $\CN=2$ Maldacena-Nu\~nez backgrounds \MaldacenaMW. One of the main results of \BeniniMZ\ is that when SUSY is broken to $\CN=1$ by integrating out the adjoint chiral superfield, the resulting theories are dual to the $\CN=1$ Maldacena-Nu\~nez solutions. This result is also hinted at by the ratio of the central charges in the UV and IR, as discussed in \TachikawaTT. 

In this work, we continue the program of \BeniniMZ\ by studying further $\CN=1$ deformations of $T_N$ theories. In particular, we consider theories which are coupled to $\CN=1$ vector multiplets, but do not have the superpotentials that result from integrating out the adjoint chiral superfields. This is equivalent to taking a limit where the mass of the adjoint is large. In such cases, there are more global symmetries than in the theories of \BeniniMZ, and we will often need to use $a$-maximization to determine the superconformal R-symmetry. We note that if the resulting theories are good SCFTs, they should then be thought of as belonging to the family of SCFTs that come from six-dimensional $(2,0)$-theories descending from M5-branes wrapping a Riemann surface.

The theories we consider in this paper involve $T_N$ theories with gauged global symmetries connected by bifundamental $\CN=1$ chiral superfields. The requisite building blocks are illustrated in Figure 1. For simplicity, in this work all of our bifundamentals will have the same matter content as an $\CN=2$ hypermultiplet. We will additionally consider matter fields which are adjoints, since these will  show up in theories which are the putative endpoints of certain RG flows. 

\topinsert
\bigskip
\centerline{\hskip1cm\epsfxsize=.5\hsize\epsffile{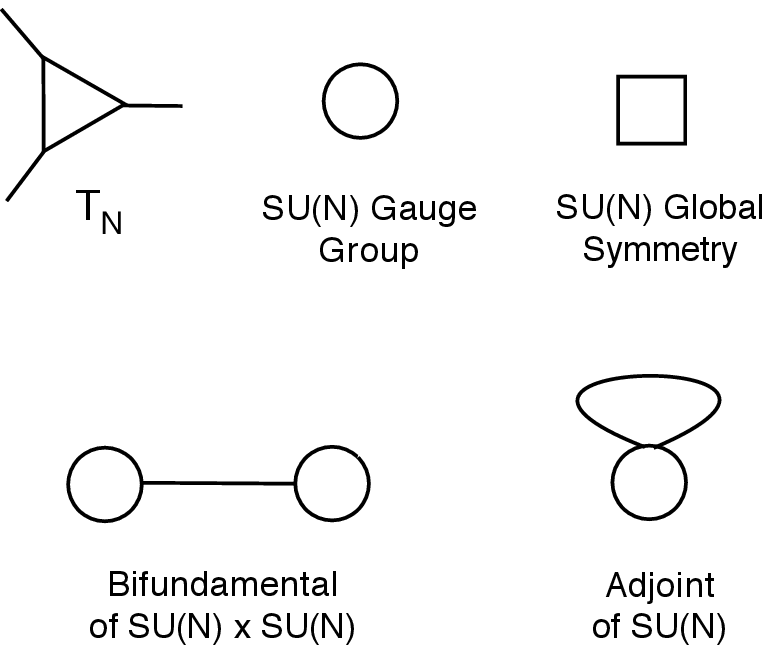}}
\centerline{\ninepoint\sl \baselineskip=8pt {\bf Figure 1:}
The basic building blocks of our theories.}
\bigskip
\endinsert

Our main question is whether the the theories we consider are consistent, dynamically realized SCFTs. Of course, in the absence of some kind of duality, it is very difficult to determine whether or not these theories actually achieve an interacting conformal phase. However, we provide a series of checks that some of the theories we consider are in fact new $\CN=1$ SCFTs without Lagrangian descriptions. We use two main criteria: First, we check that the gauge-invariant operators have dimensions $\Delta$ that obey the unitarity bound $\Delta \geq 1$. Second, we check that when we deform theories by a relevant operator, or give a vev to a field, the resulting flow does not violate the $a$-theorem (recently proved in \KomargodskiVJ). 

We find that when we connect $T_N$ blocks with only hypermultiplets and do not turn on a superpotential, these theories appear to be good SCFTs, in the sense described above. In particular, there are no unitarity bound violations, and when we turn off the superpotential of \BeniniMZ, the resulting flows are consistent with the $a$-theorem. In contrast, it appears that whenever we add adjoints, the resulting theories do not appear to be consistent SCFTs, since we see flows which are inconsistent with the $a$-theorem.

The rest of this paper is structured as follows. In Section 2, we review some relevant facts about $\CN=1$ and $\CN=2$ SCFTs and briefly describe our setup. Section 3 describes our flagship family of theories, in which $T_N$ blocks are connected via hypermultplets. We note that some of these theories require $a$-maximization, which we perform. In Section 4 we describe some other possible $\CN=1$ setups involving $T_N$ blocks; these theories will show up when we analyze the flows from the theories of Section 3. In Section 5 we describe these flows, note that some of them appear to violate the $a$-theorem, and discuss possible resolutions. Finally in Section 6, we briefly conclude.

\newsec{Review and Setup}

In this section we review the basics of $T_N$ theories from both an $\CN=1$ and $\CN=2$ perspective, and describe the techniques necessary to explore the $\CN=1$ deformations of these theories.

\subsec{Review of $T_N$ Technology}

$T_N$ theories were discovered by Gaiotto \GaiottoWE\ via S-duality. They describe the low-energy four-dimensional $\CN=2$ theories coming from $N$ M5-branes wrapping a thrice-punctured Riemann surface. For $N \geq 3$, these theories have no known weakly coupled Lagrangian descriptions. Although such a fundamental description is lacking, we still have access to a relatively large amount of information, such as global symmetries and operator dimensions. For our purposes here, this information will turn out to be exactly what we need.

The global symmetries of a $T_N$ theory are $SU(2)_R \times U(1)_R \times SU(N)^3$. The Coulomb branch is parameterized by operators $u_k^{(i)}$, with $k = 3,...,N$ and  $i=1,...,(k-2)$. These operators have scaling dimension $\Delta(u_k^{(i)})=k$. The Higgs branch is described by dimension-two operators $\mu_\alpha$, for $\alpha = 1,...,3.$ Each $\mu_\alpha$ transforms in the adjoint representation of one of the $SU(N)$ global symmetries. There are also dimension $(N-1)$ operators $Q_{ijk}$ and $\widetilde {Q}^{ijk}$ that are in the trifundamental\foot{For an interesting recent discussion of $\CN=2$ theories with trifundamentals, see \TachikawaEA.} $({\bf N},{\bf N},{\bf N})$ and $(\overline {\bf N}, \overline {\bf  N}, \overline {\bf N})$ of  $SU(N)^3$. 

It is convenient to single out the two abelian symmetries inside the $SU(2) \times U(1)$ R-symmetry. We will use $I_3$ to denote the $U(1)$ generated by $T^3  \subset SU(2)$, and the other $U(1)$ by $R_{\CN=2}$. To fix notation, recall that for free vectors and hypers these symmetries have charges as follows:
\eqn\charges{
\matrix{R_{\CN=2} \setminus I_3 & {1 \over 2} & 0 & -{1 \over 2} \cr
0 & & A_\mu & \cr
1 & \lambda & & \lambda' \cr
2 & & \phi & } \qquad \qquad
\matrix{R_{\CN=2} \setminus I_3 & {1 \over 2} & 0 & -{1 \over 2} \cr
-1 & & \psi & \cr
0 & Q & & \widetilde Q^\dagger \cr
1 & & \widetilde \psi^\dagger & }
}
For a $T_N$, the nonzero 't Hooft anomalies for these two global symmetries are  \GaiottoGZ:
\eqn\tnanom{\eqalign{
\Tr_{T_N} R_{\CN=2}^3 &= \Tr_{T_N} R_{\CN=2} = 2 + N - 3N^2 \cr
\Tr_{T_N} R_{\CN=2} I_3^2 & = {1 \over 12} \left ( 6 - N - 9 N^2 + 4N^3 \right ).
}}
Similarly, we can compute the central charges $a$ and $c$:
\eqn\aandc{
\eqalign{ a_{T_N} &=  {N^3 \over 6} - {5 N^2 \over 16} -{N \over 16} +{5 \over 24} \cr
c_{T_N} &= {N^3 \over 6} - {N^2 \over 4} - {N \over 12} + {1 \over 6}.
}}
We will also need to use the fact that if we gauge one of the $SU(N)$ global symmetries of the $T_N$,
\eqn\tnr{
\Tr_{T_N} R_{\CN=2} T^a T^b = -N \delta^{ab},
}
where $T^{a}$ is a generator of the gauged symmetry. For a derivation, see  \BeniniMZ\ or \GaiottoGZ.

The purpose of the present paper is to extend the program of \BeniniMZ\ by studying the $\CN=1$ properties of $T_N$ theories. To this end, we will make frequent use of various linear combinations of $R_{\CN=2}$ and $I_3$. One particularly useful choice is
\eqn\lincomb{\eqalign{
R_{\CN=1} &= {1 \over 3} R_{\CN=2} + {4 \over 3} I_3 \cr
J &= R_{\CN=2} - 2 I_3.
}}
By checking the charge assignments in \charges, one can see that $J$ is a non-R symmetry. $R_{\CN=1}$ is the superconformal R-symmetry for a free $\CN=1$ theory.

\subsec{Breaking to $\CN=1$}

In this paper, we will be working with $T_N$'s in theories which have only $\CN=1$ SUSY. The theories we consider will have at least one $T_N$ for which we gauge part of the global symmetry group, and take the gauge boson to be part of an $\CN=1$ vector multiplet. One way to arrive at such a theory is to start with an $\CN=2$ vector mutiplet, and give a mass to the associated $\CN=1$ adjoint chiral superfield $\Phi$ via a superpotential $W = m \Tr \Phi^2$. The R-symmetry preserved by this deformation is 
\eqn\rzerois{
R_0 = R_{\CN=1} + {1 \over 6} J = {1 \over 2} R_{\CN=2} + I_3.
}
In the case where the vector mutiplet is coupled to a hypermultiplet $Q \oplus \widetilde Q$, $R_0$ is the R-symmetry that gives charge 1/2 to both $Q$ and $\widetilde Q$, making the superpotential $W = {1 \over m} (Q \widetilde Q)^2$ marginal.

The main question we wish to answer here is whether or not there exist good (consistent, dynamically realized) SCFTs in the absence of such a superpotential term. One way to achieve this would be to take $m$ large, in which case the superpotential vanishes. Alternately, one could just consider the theory as coupled only to $\CN=1$ vector multiplets from the start. For our purposes here we will simply turn off this superpotential and then check that the theory satisfies a reasonable set of consistency conditions. 

In practice, turning off this superpotential means that the theory can have additional $U(1)$ global symmetries. For example, in a theory with only vector and hypermultiplets, the $U(1)$ denoted by $J$ in \lincomb\ is no longer explicitly broken by the superpotential. This is just the axial $U(1)$ that comes with any $Q \oplus \widetilde Q$. Such a symmetry is forbidden by $\CN=2$ SUSY, but is of course allowed in $\CN=1$ theories. When we consider $T_N$ blocks in $\CN=1$ theories, each $T_N$ will similarly come equipped with an ``axial" $J$, whose action on the $T_N$ is given by the linear combination of $U(1)$'s in \lincomb.

\subsec{$a$-maximization}

Since we will generically not be able to use symmetries to uniquely determine the superconformal R-symmetry, we will need to use $a$-maximization \IW, which we now review. In general, any putative R-symmetry $R_{trial}$ will have the form
\eqn\rtrial{
R_{trial} = R_0 + \sum_I s_I F_I,
}
where $I$ runs over all $U(1)$ global symmetries $F_I$, and the $s_I$ are as yet undetermined coefficients. These $s_I$ are uniquely specified by maximizing the function
\eqn\atrial{
a_{trial}(s_I) = 3 \Tr R_{trial}^3 - \Tr R_{trial}.
}
As first noted in \IW, global symmetries $F_I$ which have $\Tr F_I = 0$ will never mix with the R-symmetry; this follows from the extremum condition $9 \Tr R^2 F_I = \Tr F_I$, which means that when $\Tr F_I = 0$ the R-symmetry can always be taken to commute with $F_I$. The most prominent example of such a symmetry is the baryonic $U(1)$ in SQCD. Thus we will often refer to such $U(1)$'s as ``baryonic." 

We now record some formulae that will be useful for our computations. The contribution to $a_{trial}$ from each $T_N$ can be computed by
writing $R_0$ and $J$ in terms of $R_{\CN=2}$ and $I_3$ as in \lincomb\ and \rzerois. Taking the coefficient of $J$ to be $s$, this contribution is
\eqn\atn{
a_{trial}^{T_N} = \left(s + {1 \over 2}\right)\left ( 3s^2 +3s -{1 \over 4}\right) \Tr_{T_N}R_{\CN=2}^3 + 9\left(s + {1 \over 2}\right)(1-2s)^2\, \Tr_{T_N}R_{\CN=2}I_3^2.
}
In this expression, we have used $\Tr_{T_N} R_{\CN=2} = \Tr_{T_N} R_{\CN=2}^3$. The two 't Hooft anomalies in \atn\ are given in \tnanom. 

For reference, we additionally note that in $\CN=1$ theories, a hypermultiplet comes with both an axial and a baryonic $U(1)$. For our bifundamental hypermultiplets, the contribution of the axial $U(1)$ to $a_{trial}$ is 
\eqn\ahyp{
a_{trial}^{hyper} = 2N^2 \left [ 3\left (s-{1\over 2}\right )^3- \left (s- {1 \over 2}\right ) \right ],
}
where $s$ is the coefficient of $F$ in \rtrial. This formula, along with \atn, will be essentially all we need in order to do computations in our $\CN=1$ theories.

\subsec{Building new $\CN=1$ theories}

In this paper we will construct theories with three different elements: $T_N$ blocks, $\CN=1$ vector mutiplets, and matter chiral superfields. We can depict any such theory by a diagram whose elements are shown in Figure 1. One difference with \BeniniMZ\ is that in the present work, we will occasionally include adjoint chiral superfields. We will treat these adjoints as independent $\CN=1$ superfields, and not as part of an $\CN=2$ vector multiplet. This situation will occur naturally in some of the RG flows we study.

The easiest case to consider is one in which all gauge and global symmetry groups are $SU(N)$. We restrict ourselves to such theories in this work. The theories considered here have several types of matter. First, there are bifundamental hypermultiplets $Q \oplus \widetilde Q$, which transform in the $({\bf N}, \overline {\bf N}) \oplus (\overline {\bf N}, {\bf N})$ of an $SU(N) \times SU(N)$ gauge group. Additionally, we can have adjoint, fundamental, or antifundamental fields of a given $SU(N)$ gauge group. Finally, we allow $T_N$'s, as described above.

One particular subclass of such theories is given by a chain of hypermultiplets with a $T_N$ on each end, as in Figure 2. This theory is uniquely specified by the number of hypermultiplets $\ell$, and has no superpotential. For convenience, we will denote such theories by the name $\CS_\ell$. They will be the flagship examples of the present work. 

\topinsert
\bigskip
\centerline{\epsfxsize=.5\hsize\epsffile{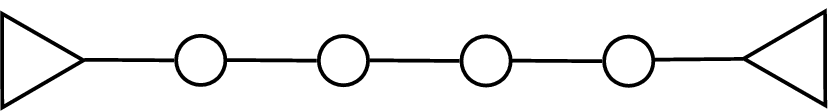}}
\centerline{\ninepoint\sl \baselineskip=8pt {\bf Figure 2:}
An example of an $\CS_\ell$ theory with $\ell = 3$. The circles are $\CN=1$ vector multiplets.}
\bigskip
\endinsert

We are now ready to begin checking whether or not $\CS_\ell$ theories are good SCFTs.

\newsec{New $\CN=1$ Theories}

In this section, we provide some consistency checks  that $\CS_\ell$ theories are likely to be interacting $\CN=1$ SCFTs. We begin by doing some simple examples before moving on to the more general case.

\subsec{$\CS_0$}


Consider a theory with two $T_N$'s and one vector multiplet. As per our classification, we call this quiver ${\cal S}_0$. This theory has an anomaly-free R-symmetry $R_0$, given in \rzerois. Additionally, each $T_N$ has a $U(1)$ global symmetry $J_{1,2}$, for which the anomaly-free linear combination is
\eqn\jszero{
\CF = J_1 - J_2.
}
This global $U(1)$ is baryonic, since $\Tr \, \CF = 0$. As discussed above, such a global symmetry will never mix with the R-symmetry. Thus the superconformal R-symmetry here is simply $R_0$, and it is straightforward to check that no gauge-invariant operators violate the unitarity bound.

\subsec{$\CS_1$}

Now consider a theory with two $T_N$'s, two vectors, and one hyper. In addition to $J_1$ and $J_2$, there is an axial $U(1)$, $F$, acting only on $Q\oplus \widetilde{Q}$. We adopt the normalization $F(Q)=F(\widetilde Q)=1$ (note that $Q$ and $\widetilde Q$ must have the same charge because of charge conjugation symmetry).  The hypermultiplet also comes with an anomaly-free baryonic $U(1)$, which we ignore since it does not mix with $R_0$. There is an anomaly-free global symmetry given by 
\eqn\jjf{
{\cal F} \equiv J_1 + F + J_2.
}
Since $\Tr {\cal F} \, \neq 0$, $\CF$ can mix with $R_0$.  Therefore there is a one-parameter family of potential R-symmetries, given by
\eqn\rtriallone{
R_{trial} (\alpha) = R_0 + \alpha {\cal F}.
} 
We fix the superconformal R-symmetry by $a$-maximization.  Using \atn\ and \ahyp,  we find that $a_{trial}$ is maximized at 
\eqn\alphaans{
\hat \alpha(N) = { - 4 N + 3 N^2 + 4N^3 - 2 \sqrt{16 - 16N - 52 N^2 + 48 N^3 + 33 N^4 - 44 N^5 + 16 N^6} \over 6(8-11 N^2 + 4N^3)}.
}
This function is negative and monotonically decreasing for $N \geq 2$, and approaches $-1/6$ at large $N$. See Figure 3.

\bigskip
\centerline{\epsfxsize=.5\hsize\epsfbox{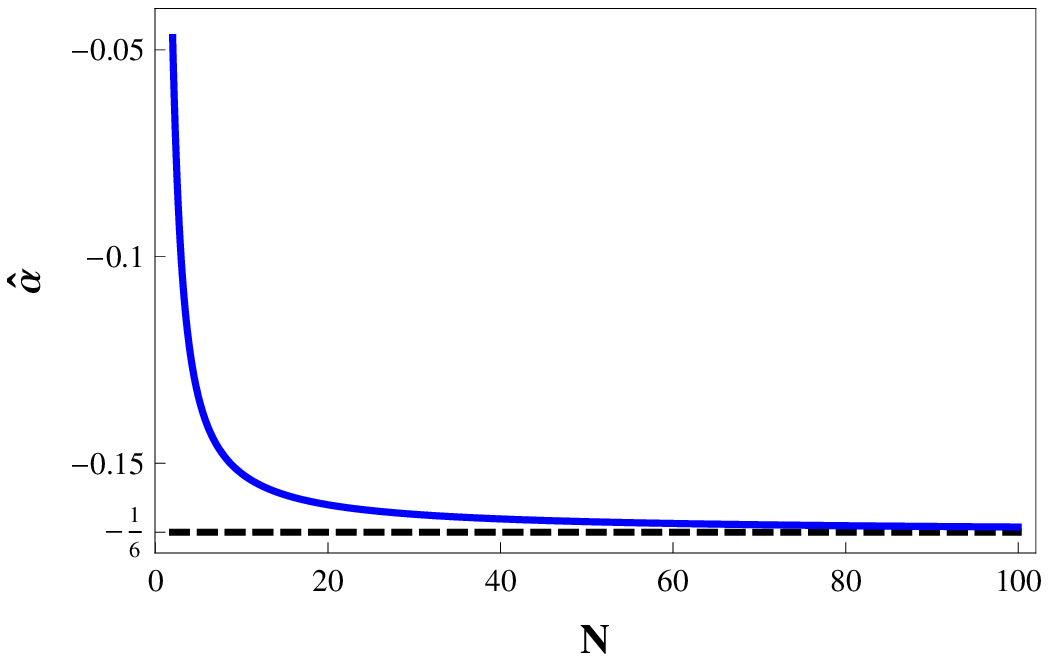}}
\centerline{\ninepoint\sl \baselineskip=8pt {\bf Figure 3:}
{\sl $\hat{\alpha}(N)$ for $\CS_1$. The dotted line is the asymptotic value $-1/6$.}}
\bigskip

$\CS_1$ has various gauge-invariant chiral operators constructed from $Q, \widetilde Q,$ and $\mu$. The R-charges of $Q, \widetilde Q,$ and $\mu$ are
\eqn\rcqmu{
R(Q) = R(\widetilde Q) = {1\over 2} + \hat \alpha, \;\;\; R(\mu) = 1-2\hat \alpha. } Because we have only gauged two of the global $SU(N)$'s, four $\mu_i$ remain gauge-invariant. We will denote the remaining two $\mu$ operators, which transform in the adjoint of a gauged $SU(N)$, by $\mu_{L,R}$.
Thus, the gauge-invariant operators and their superconformal R-charges are schematically
\eqn\rcginv{
\matrix{ {\rm operator} & R(\hat \alpha) \cr \mu_i & 1-2\hat{\alpha} \cr Q \mu_{L,R} \widetilde{Q} & 2 \cr  Q\widetilde{Q} & 1+ 2\hat{\alpha} \cr Q\widetilde{Q}Q\widetilde{Q} & 2+ 4\hat{\alpha}.}}
Since $\hat \alpha \rightarrow -1/6$ monotonically, no operators violate the unitarity bound $R \geq 2/3$. Thus this theory appears to satisfy one of the necessary criteria for the existence of an SCFT.
  
We note that at large $N$, the R-charge of $Q \widetilde Q$ approaches 2/3, so this operator becomes free and decouples from the theory. This may prove problematic when trying to find a gravity dual for these theories, if one exists.

\subsec{General $\CS_\ell$}

Now we consider the general $\CS_\ell$ theory, which has $\ell$ bifundamental hypers.  We denote the hypermultiplet connecting the $i^{\rm th}$ and $(i+1)^{\rm th}$ node by $Q_i \oplus \widetilde{Q}_i$, where we take the first node to be $i=0$. Each hyper has an axial $U(1)$ symmetry $F_i$, with charges $F_i (Q_j) = F_i (\widetilde Q_j) = \delta_{ij}.$ The unique anomaly-free non-R $U(1)$ is 
\eqn\uonel{
{\cal F}_\ell \equiv J_1 + (-1)^{\ell-1} J_2 + F_o -F_e} 
where
\eqn\fefo{
F_e = \sum_{i=1}^{\lfloor {\ell \over 2} \rfloor}F_{2i}, \;\;\; F_o = \sum_{i=1}^{\lfloor {\ell + 1 \over 2} \rfloor} F_{2i-1}.}

Under the refection symmetry of the quiver, ${\cal F}_\ell \to (-1)^{\ell+1} {\cal F}_\ell$.  When $\ell$ is even, $\Tr {\cal F}_\ell = 0$, and this $U(1)$ will not mix with the R-symmetry. Thus, for even $\ell$, the superconformal R-symmetry is $R_0$. When $\ell$ is odd, the candidate R-symmetry is 
$R_{trial} = R_0 + \alpha {\cal F}_\ell.$ $a$-maximization yields 
\eqn\alphaansl{      
\hat \alpha(\ell,N)  = {A - \sqrt{B} \over C},
}
where
\eqn\abc{\eqalign{
A &= - 4 N + 3 \ell N^2 + 4N^3 \cr
B &= 64- 64N - 208 N^2 -24(\ell-9)N^3 + 3(3\ell^2+41) N^4 +8(3\ell-25)N^5 + 64 N^6 \cr
C &=6(8-11 N^2 + 4N^3).
}}

\bigskip
\centerline{\epsfxsize=.5\hsize\epsfbox{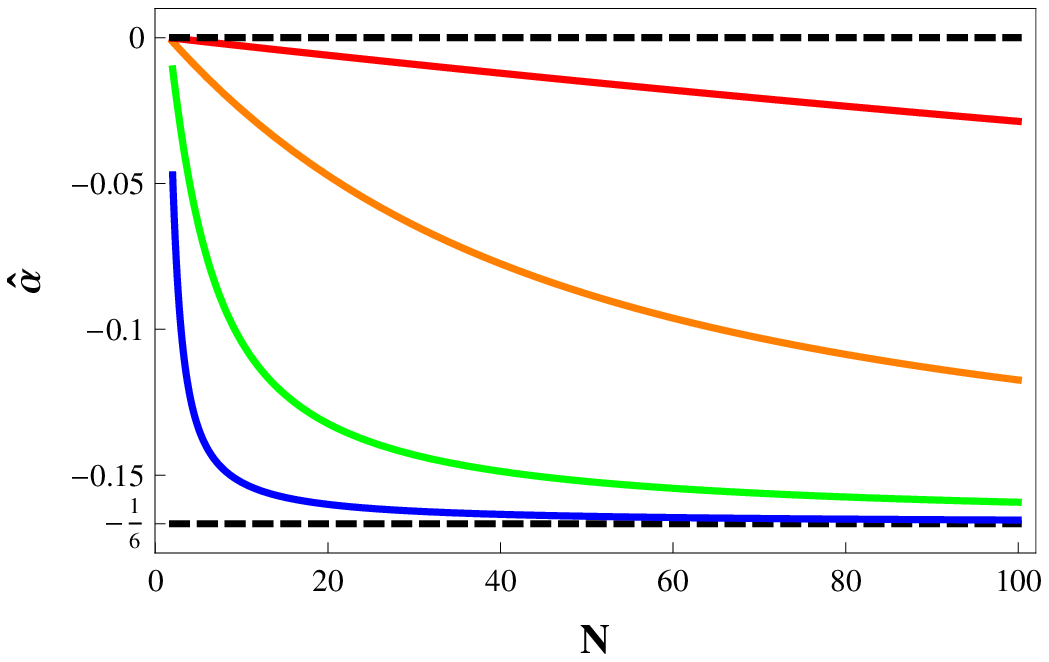}}
\centerline{\ninepoint\sl \baselineskip=8pt {\bf Figure 4:}
{\sl $\hat{\alpha}(\ell, N)$ for various $\CS_\ell$ as a function of $N$.}}
\centerline{\ninepoint\sl \baselineskip=8pt The colors correspond to $\ell =1$ (blue, bottom), 11, 101, and 1001 (red, top).}
\bigskip

We plot $\hat \alpha(\ell,N)$ in Figure 4. For all values of $\ell$ and $N$, $\hat{\alpha}$ is negative.  At fixed $\ell$, $\hat \alpha$ is a monotonically decreasing function of $N$ that asymptotes to $-1/ 6$.  At fixed $N$, $\hat{\alpha}$ monotonically increases and approaches zero. In this case, the R-charges of operators in theories with odd and even $\ell$ become the same. 

We can now compute the dimensions of all the chiral primary operators in the theory. For even $\ell$, all $R(Q) = R(\widetilde Q ) = 1/2$, and $R(\mu) = 1$ for any $Q, \widetilde Q$, and $\mu$. The only relevant operators in this theory are $Q \widetilde Q$. The other gauge-invariant operators, $\mu Q \widetilde Q$ and $Q \widetilde Q Q \widetilde Q$, are marginal. 

\bigskip
\centerline{\epsfxsize=.6\hsize\epsfbox{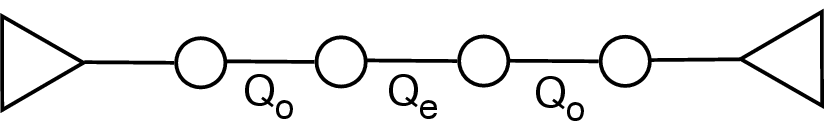}}
\centerline{\ninepoint\sl \baselineskip=8pt {\bf Figure 5:}
{\sl $\CS_3$, with $Q_o$ and $Q_e$ marked.}}
\bigskip

When $\ell$ is odd,  the dimension of a hypermultiplet depends on its position in the quiver. This follows from the different signs of $F_o$ and $F_e$ in \uonel. We will denote bifundamentals which are to the left of odd numbered nodes by $Q_{o}, \widetilde Q_o$, and the other bifundamentals by $Q_{e},\widetilde Q_e$. See Figure 5. From \uonel, we see that these fields have R-charges
\eqn\evenoddq{
R(Q_o) = R(\widetilde Q_o) = {1 \over 2} + \hat \alpha, \qquad R(Q_e) = R(\widetilde Q_e) = {1 \over 2} - \hat \alpha.
}
The simplest gauge-invariant operators we can form from these operators are quadratic. Such fields have R-charges
\eqn\quadops{
R( Q_o \widetilde Q_o ) = 1 + 2 \hat \alpha, \qquad R(Q_e \widetilde Q_e) = 1 - 2 \hat \alpha.
}
We can also make quartic operators, with R-charges
\eqn\quartops{
R( (Q_o \widetilde Q_o)^2 ) = 2 + 4 \hat \alpha, \qquad R( (Q_e \widetilde Q_e)^2 ) = 2 - 4 \hat \alpha, \qquad R(Q_o \widetilde Q_o Q_e \widetilde Q_e ) = 2.
}
Finally, we note that the gauge-invariant operators involving $\mu_i$ have R-charge
\eqn\muqq{
R( \mu_i Q_o \widetilde Q_o ) = 2.
}
There are no operators of the form $\mu_i Q_e \widetilde Q_e$ in theories with odd $\ell$.

At fixed $\ell$, $\hat \alpha$ is a monotonic function that approaches $-1/6$ at large $N$. In this case, it is easy to see that only the quadratic operators and $(Q_o \widetilde Q_o)^2$ can be relevant deformations of the theory. When we fix $N$ but take $\ell$ large, $\hat \alpha \rightarrow 0$. Thus the quadratic operators are relevant and the quartic operators are all marginal. Note that for both even and odd $\ell$, all gauge-invariant operators respect the unitarity bound for any values of $\ell$ and $N$.

\newsec{Other $\CN=1$ theories}

We now move on to consider other $\CN=1$ theories which are similar to the $\CS_\ell$ theories constructed above.

\subsec{Theories with one $T_N$.}

\bigskip
\centerline{\epsfxsize=.5\hsize\epsfbox{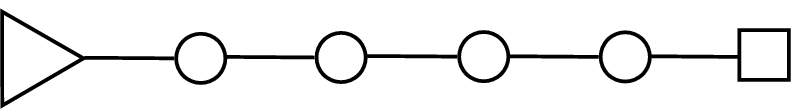}}
\centerline{\ninepoint\sl \baselineskip=8pt {\bf Figure 6:}
{\sl The theory $\CS^{\square}_3$.}}
\bigskip

In this section, we study $\CS_\ell$ theories where the right $T_N$ has been replaced by $N$ fundamental hypermultiplets, as in Figure 6. We will name these theories $\CS^{\square}_\ell$.  Since the $T_N$'s and $N$ hypermultiplets have the same contribution to the gauge anomaly, $R_0$ is still anomaly-free.  The total number of hypermultiplets is $\ell +1$, where $\ell$ of the hypers are bifundamental. This theory has an anomaly-free non-R $U(1)$ global symmetry
\eqn\uonef{
{\cal F}_\ell = J - \sum_i^\ell (-1)^i F_i + (-1)^\ell F_f}
where $J$ acts on the $T_N$ and $F_f$ acts on the fundamental hypermultiplets with charge 1.  The trial R-symmetry is 
\eqn\rtrialf{
R_{trial} = R_0 + \alpha {\cal F}_\ell.}

Once again, we can determine $\hat \alpha$ by $a$-maximization. We plot the result in Figure 7.  At fixed $\ell$, $\hat{\alpha}$ is a monotonically decreasing function of $N$ that approaches $-{1\over 6}$.  At $N=2$, it vanishes when $\ell$ is even.  This follows from the fact that $T_2$ is a theory of two hypermultiplets.  In this case, the quiver has a reflection symmetry which switches the $T_2$ on the left with the two fundamentals on the right.  When $\ell$ is even, ${\cal F}_\ell$ is odd under reflection while $R_0$ is even.  Thus the two $U(1)$'s cannot mix.  In general, $\hat{\alpha}$ is bounded above by zero.  The behavior of the gauge-invariant operators constructed from the hypermultiplets is the same as in the $\CS_\ell$ theories.  As a consequence, they do not violate the unitarity bound.

\bigskip
\centerline{\epsfxsize=.5\hsize\epsfbox{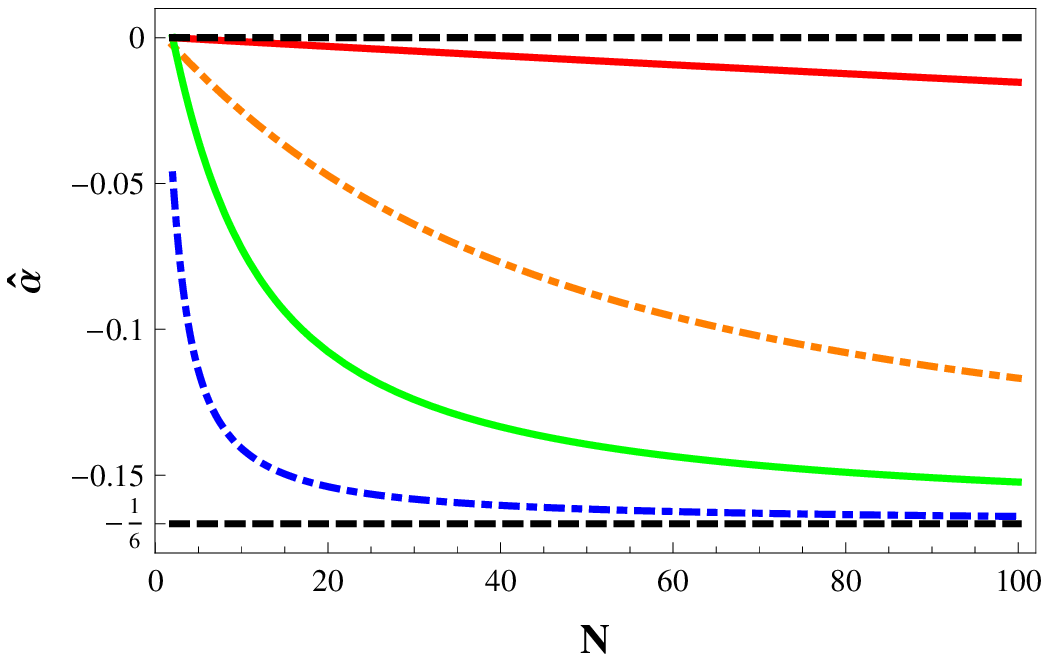}}
\centerline{\ninepoint\sl \baselineskip=8pt {\bf Figure 7:}
{\sl $\hat{\alpha}(\ell,N)$ for $\CS^{\square}_\ell$ at $\ell =1$ (dashed blue, bottom), $10,51,1000$ (solid red, top).}}
\bigskip

\subsec{Ending with adjoints}

\bigskip
\centerline{\epsfxsize=.5\hsize\epsfbox{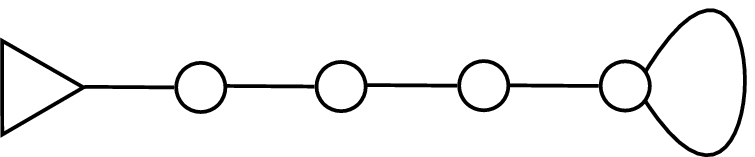}}
\centerline{\ninepoint\sl \baselineskip=8pt {\bf Figure 8:}
{\sl The theory $\CS^{\circ}_3$.}}
\bigskip

We now add adjoint matter to the $\CS_\ell$ theories. The simplest thing we can do is substitute the $T_N$ at the right end of the quiver with a chiral superfield $X$ in the adjoint of $SU(N)$, as in Figure 8. We will call these theories $\CS^\circ_\ell$.  $R_0$ is anomaly-free when $R_0(X)={1\over 2}$.  Additionally, $X$ comes with a $U(1)$, $F_a$, which we normalize as $F_a(X)=1$.  The theory then has an anomaly-free global $U(1)$ given by
\eqn\fa{
{\cal F} = J_1 - \sum_{i=1}^{\ell} (-1)^iF_i + (-1)^\ell F_a.}
The trial R-symmetry can be written as
\eqn\rtriala{
R_{trial} = R_0 + \alpha {\cal F},}
so the charge of $X$ and its contribution to the trial central charge are
\eqn\rtxatx{\eqalign{
R_{trial}(X) &= {1\over 2} + (-1)^\ell \alpha \cr
a_{trial}^{X,\circ}&= (N^2 -1) \left[3\left(-{1 \over 2} + (-1)^\ell \alpha \right)^3 - \left(-{1 \over 2} + (-1)^\ell \alpha \right) \right].}}

Maximizing $a$ yields an answer very similar to that of the $\CS_\ell$ theories, although for simplicity we do not record the full answer here. Instead, we note that $\hat{\alpha}$ is a monotonically decreasing function of $N$ that approaches $-{1 \over 6}$ in $N \gg \ell$ limit.  When $\ell \gg N$, $\hat \alpha$ approaches zero; this is consistent with the contribution of the hypermultiplets dominating over those of the $T_N$.  In Figure 9, we plot $\hat{\alpha}$ as a function $N$ at different values of $\ell$.

\bigskip
\centerline{\epsfxsize=.5\hsize\epsfbox{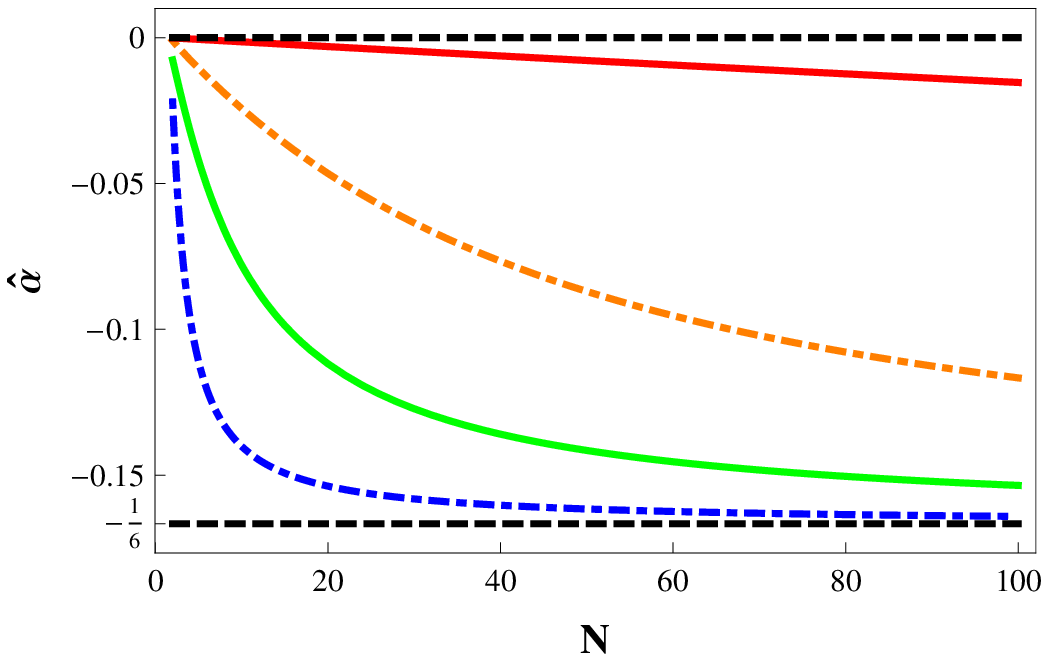}}
\centerline{\ninepoint\sl \baselineskip=8pt {\bf Figure 9:}
{\sl $\hat{\alpha}(\ell,N)$ for $\CS_\ell^\circ$ at $\ell =1$ (dashed blue, bottom), $10,51,1000$ (solid red, top).}}
\bigskip

The gauge-invariant operators in this theory do not violate the unitarity bound. This is easy to see for operators not involving $X$, since $\hat \alpha$ is again a monotonic function with the same asymptotic behavior as in the $\CS_\ell$ theories. The lowest-dimension operators involving $X$ are $\Tr X^2$ and $Q_\ell X \widetilde{Q}_\ell$, which have R-charges
\eqn\rxqq{
R(X^2)= 1+ 2(-1)^\ell \hat{\alpha}, \;\;\; R(Q_\ell X \widetilde{Q}_{\ell}) = {3 \over 2} - (-1)^\ell \hat{\alpha}.}
Since $-1/6 < \hat{\alpha} < 0$, these operators are always above the unitarity bound, and relevant. 

\subsec{Inserting adjoints in the middle}

Now we consider a different theory, with the adjoint chiral field $X$ attached to the $k^{th}$ $SU(N)$ gauge group of the $\CS_\ell$ theories (we let $k=0$ be the first node, the one attached to the left $T_N$). We will name these theories $\CS_{\ell,k}$. See Figure 10. To preserve the R-anomaly, $R_0(X) = 1$. Here the adjoint leads to an additional anomaly-free global $U(1)$, since we have increased the number of fields but not the number of constraints on their charges. Taking $F_a(X) =1$, a convenient choice for these global $U(1)$'s is
\eqn\uonebasis{\eqalign{
{\cal F}^L &=  J_1 - \sum_{i=1}^{k-1} (-1)^iF_i - (-1)^k F_a \cr
{\cal F}^R &=  J_2 + \sum_{i=k}^{\ell} (-1)^{\ell+i}F_i - (-1)^{\ell+k} F_a. }}
Note that ${\cal F}_\ell$ in \uonel\ can be written ${\cal F}_\ell = {\cal F}^L -(-1)^\ell {\cal F}^R$.

\bigskip
\centerline{\epsfxsize=.5\hsize\epsfbox{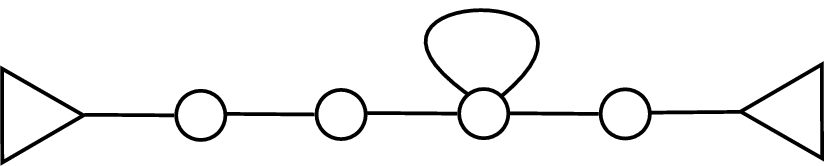}}
\centerline{\ninepoint\sl \baselineskip=8pt {\bf Figure 10:}
{\sl The theory $\CS_{3,2}$.}}
\bigskip

From the trial R-symmetry
\eqn\rtrialad{
R_{trial} = R_0 + \alpha_L {\cal F}^L + \alpha_R {\cal F}^R,} 
we observe that we need to $a$-maximize over two variables.  The charge of $X$ and its contribution to $a_{trial}$ are
\eqn\rtX{\eqalign{
R_{trial}(X) &= 1-(-1)^{k}(\alpha_L+(-1)^\ell \alpha_R), \cr a_{trial}^{X,\ell,k} &=-(-1)^k(N^2-1) \left ( 3 (\alpha_L+(-1)^\ell \alpha_R)^3 - (\alpha_L+(-1)^\ell \alpha_R) \right ).}}

After $a$-maximization, we find the following:

\lfm{1.} At fixed $N$, the value of $(\hat{\alpha}_L,\hat{\alpha}_R)$ at $(l,k)$ is equal to $(\hat{\alpha}_R,\hat{\alpha}_L)$ at $(l,l-k+2)$.  This follows from the reflection symmetry of the quiver, which switches the two $T_N$'s and maps the $i^{th}$ gauge group to $\ell-i+2$.  The adjoint which was attached to node $k$ moves to node $l-k+2$.  Under this reflection, $\hat{\alpha}_L\leftrightarrow \hat{\alpha}_R$.
\lfm{2.} $\hat{\alpha}_{L,R}$ are monotonically decreasing functions of $N$ that approach $-{1 \over 6}$ in the $N \gg \ell$ limit.  They are bounded above by zero.  The R-charge of $X$ ranges from two-thirds to four-thirds.  The former occurs in the large $N$ limit of the quiver with odd values of $k$ but even values of $\ell$, whereas the latter occurs in the large $N$ limit of the quivers with even values of $k$ and $\ell$. 
 \lfm{3.} In the $\ell \gg N$ limit, we observe various phenomena depending on whether $k$ is small or large.  When $k \ll \ell$, the pair $(\hat{\alpha}_L,\hat{\alpha}_R)$ approaches $(-{1 \over 6},0)$ when $N$ is subsequently taken large.  The left side behaves as the theory in Figure 8 with small $\ell$ while the right side behaves as this theory with $\ell \gg N$.  In the case when $k \sim {\ell \over 2}$, $(\hat{\alpha}_L,\hat{\alpha}_R)$ approaches $(0,0)$ since each side is like the theory in Figure 8 with $\ell \gg N$.
\lfm{4.} For all values of $N,l,$ and $k$, the R-charge of $X$ is above the unitarity bound.  This implies that all gauge-invariant operators constructed from $X$ are above the unitarity bound.  Since the operators from the hypermultiplets are also well behaved, this theory obeys the unitary bound.  
However, in the next section, we will see that these theories appear as the natural endpoints of flows which violate the $a$-theorem. This suggests that these adjoint theories do not exist as interacting SCFTs.

\newsec{Flows}

In this section we describe some possible flows between our new $\CN=1$ theories. We consider flows driven both by deformations by relevant operators and by Higgsing, focusing in particular on the ${\cal S}_\ell$ theories. We begin by considering linear deformations of these theories, and then proceed upwards to the quartics.

\subsec{Linear Deformations}

The only possible gauge-invariant linear deformation of an ${\cal S}_\ell$ theory is $\mu_i$, which transforms in the adjoint of the remaining global $SU(N)$ groups. Since $\mu$ is a composite field (albeit one whose description in terms of fundamental fields we don't know), this deformation does not break SUSY. When $\ell$ is even, $R(\mu)=2$ but when $\ell$ is odd, $R(\mu) < 2$ and is relevant. The theory resulting from this deformation is mysterious, and we do not at present have a good understanding of the endpoint of this flow.

\subsec{Quadratic Deformations}

The mesons $Q_{o} \widetilde{Q}_{o}$ and $Q_{e} \widetilde{Q}_{e}$ are relevant for all $\ell$ and $N$.  These operators are  mass terms for the quarks.  When they are turned on, we can integrate out the associated hypermultiplets. The resulting theory consists of two disconnected parts, each ending with an $SU(N)$ gauge group with $N_f = N_c=N$ flavors. As is well-known, this group then has a quantum constraint on the moduli space, ${\rm det} \, M - B \widetilde B = \Lambda^{2N_c}$. Even if the adjoining group with $N_f = 2N_c = 2N$ becomes strong, this holomorphic, exact constraint is valid. Thus, we expect that at a generic point on this moduli space, the mesons $M$ will get vevs and break the adjacent $SU(N)$ group to $U(1)^{N-1}$. Although the gauge group will flow free in the IR, the gauge-invariant fields will be part of a complicated sigma model, whose full quantum description is difficult to analyze.

One special point we can analyze is when $\langle M \rangle = 0$. Here, the baryons get vevs, but this has no effect on the adjacent gauge groups. When the mesons for each of the disconnected parts have zero vev, the theory splits into two disconnected ${\cal S}^\circ_\ell$ theories. 
In particular, if we integrate out the bifundamental between nodes $k-1$ and $k$, the IR theory is then ${\cal S}^\circ_{k-2} \oplus {\cal S}^\circ_{\ell-k-1}$, at this special point in moduli space.

We can now compute the difference in central charges and check agreement with the recently proven $a$-theorem \KomargodskiVJ.  In Figure 11, we plot $a_{UV} -a_{IR}$ for $\ell =100$ at various values of $k$.  This behavior is generic for even values of $\ell$.  It is easy to  understand the violation at large $N$.  In this limit, the central charge is dominated by the $T_N$, whose contribution to $a$ scales as $N^3$.  As discussed in Section 3, the superconformal R-symmetry of the ${\cal S}_{\ell}$ theories is $R_0$ when $\ell$ is even. In the $T_N$ sector of this theory, the R-symmetry is $R_0-{1 \over 6} J$.  Thus the $T_N$ contributions of ${\cal S}^\circ_{k-2} \oplus {\cal S}^\circ_{\ell-k-1}$ will always be greater than ${\cal S}_{\ell}$.  

\bigskip
\centerline{{\epsfxsize=.5\hsize\epsfbox{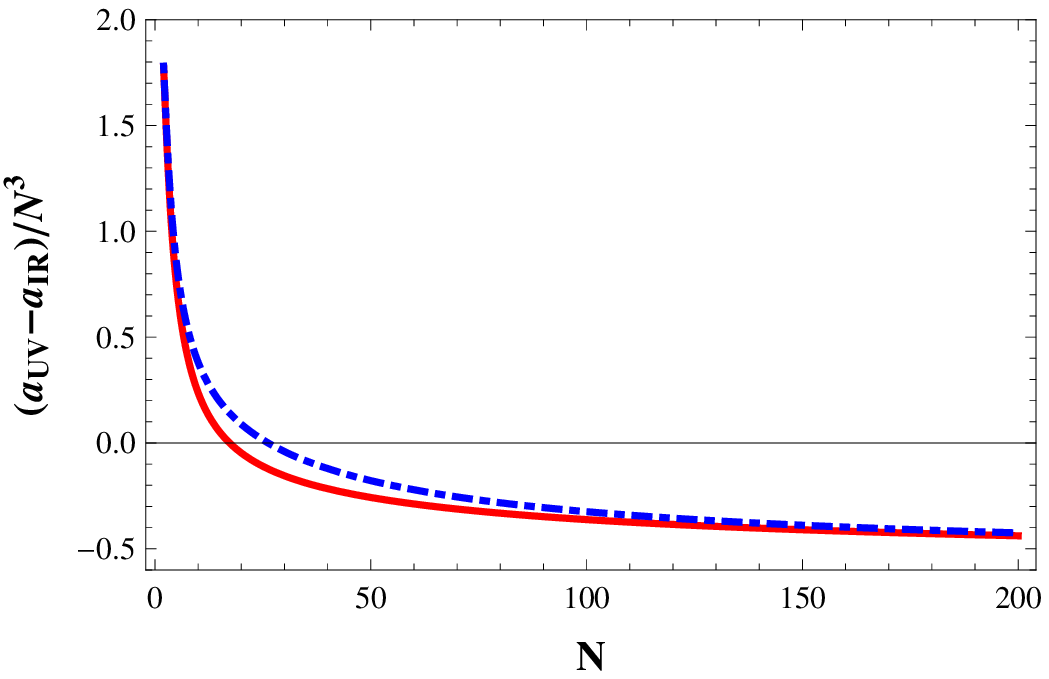}}}
\centerline{\ninepoint\sl \baselineskip=8pt {\bf Figure 11:}
{\sl Difference of central charges between ${\cal S}_{\ell}$ and ${\cal S}^\circ_{k-2} \oplus {\cal S}^\circ_{\ell-k-1}$ at $\ell=100$.}}
\centerline{\ninepoint\sl \baselineskip=8pt The colors correspond to $k=3$ (red, solid) and 53 (blue, dashed).} 
\bigskip

In Figure 12, we plot the difference in central charges for $\ell=151$ as a representative example of the case with odd $\ell$.  At large $N$, the difference is always positive.  This follows from the fact that the R-symmetry acting on the $T_N$ is always $R_0 - {1\over 6}J$ for both ${\cal S}_\ell$ and $S^\circ_{\ell}$ when $\ell$ is odd.  Thus the $T_N$ contributions cancel and any difference must come from the hypermultiplets and vector multiplets. These do not violate the $a$-theorem.  At finite $N$, we observe $a$-theorem violations when $k\ll \ell$, i.e.~when we integrate out fields close to the $T_N$.  

We can also integrate out bifundamentals in the ${\cal S}^{\square}_{\ell}$ theory and obtain an ${\cal S}^\circ_k$ theory, along with a quiver of vectors and hypermultiplets with an adjoint. We observe $a$-theorem violations at large $N$ when $k\ll \ell$. We have not included the relevant plots here, as they are quite similar to the ones already discussed. We conclude that the $\CS^{\circ}_k$ theories cannot be the endpoint of flows from either the $\CS_\ell$ or $\CS_\ell^{\square}$ theories.  Since we have no dynamical way to obtain the $\CS^{\circ}_k$ theories, it is reasonable to conclude that they do not exist.

\bigskip
\centerline{{\epsfxsize=.5\hsize\epsfbox{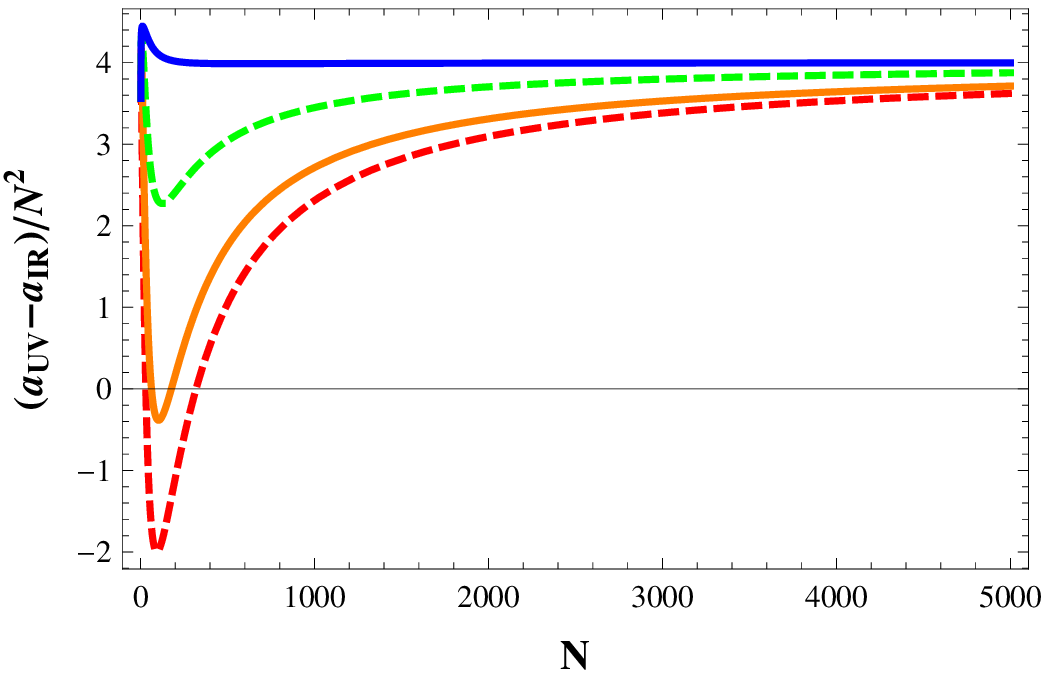}}}
\centerline{\ninepoint\sl \baselineskip=8pt {\bf Figure 12:}
{\sl Difference of central charges between ${\cal S}_{\ell}$ and ${\cal S}^\circ_{k-2} \oplus {\cal S}^\circ_{\ell-k-1}$ at $\ell=151$. }}
\centerline{\ninepoint\sl \baselineskip=8pt The color scheme is $k=3$ (dashed red, bottom), 12, 33, 60, (blue, top). } 
\bigskip

\subsec{Cubic Deformations}

There is only one possible cubic deformation, $\mu Q \widetilde Q$. This operator has R-charge 2 for both even and odd $\ell$ and thus is not a relevant deformation. Since we can add it to the theory without breaking any global symmetries, it is an exactly marginal operator and leads only to movement along the conformal manifold.

\subsec{Quartic Deformations}

For even values of $\ell$, all quartic operators are marginal. Since such operators do not break any global symmetries, they simply move the theory along the conformal manifold. We thus consider only odd values of $\ell$ for the rest of this subsection.

For odd $\ell$, the quartic operators $(Q_o \widetilde Q_o )^2$ are relevant. Adding this operator to the superpotential breaks the global $U(1)$ ${\cal F}$ but preserves $R_0$. Thus the IR R-symmetry is $R_0$. With this R-symmetry, every other possible quartic term has R-charge 2, so we can deform the theory by these operators without changing the R-symmetry. This is consistent with the fact that the deformed theory has no non-R global symmetries, so according to \GreenDA, all marginal deformations are exactly marginal. This theory is part of the conformal manifold of theories that one would get by breaking from $\CN=2$ to $\CN=1$ by giving a mass to the adjoint chiral superfiled inside the $\CN=2$ vector multiplet, i.e.~the theories studied in \BeniniMZ.  The resulting difference between the central charges is plotted in Figure 13. One can see that there is no violation of the $a$-theorem.

\bigskip
\centerline{{\epsfxsize=.5\hsize\epsfbox{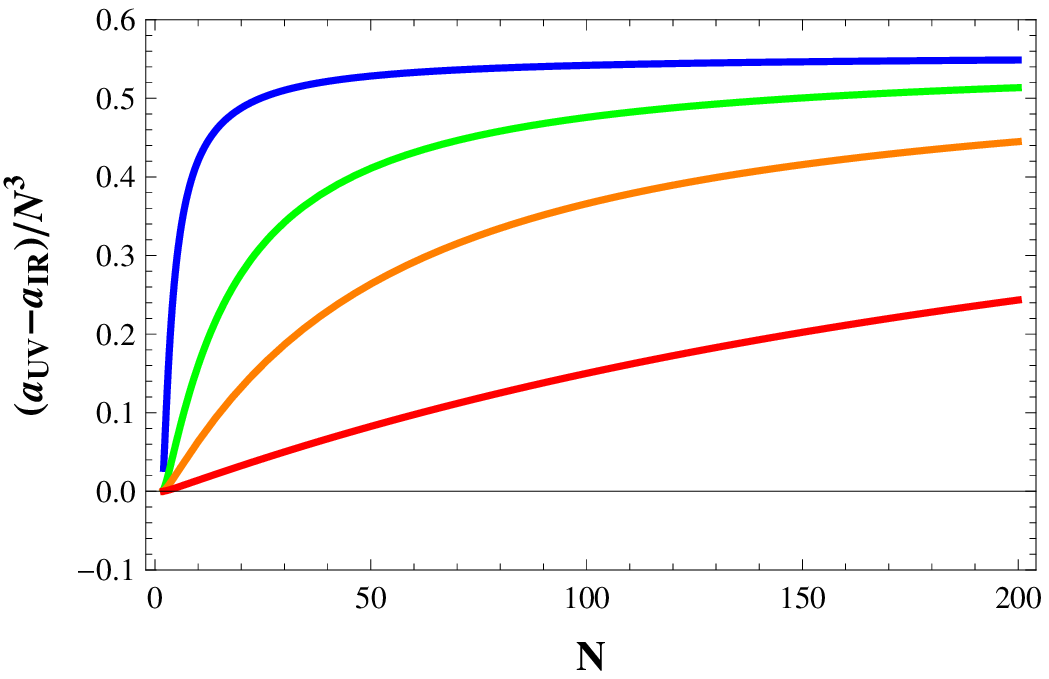}}}
\centerline{\ninepoint\sl \baselineskip=8pt {\bf Figure 13:}
{\sl Difference of central charges between ${\cal S}_{\ell}$ and ${\cal S}_{\ell}$ with $(Q_o \widetilde Q_o )^2$ turned on.}}
\centerline{\ninepoint\sl \baselineskip=8pt The colors correspond to $\ell=1$ (blue, top), 31, 101, and 501 (red, bottom).} 
\bigskip

The two other kinds of quartic operators are less interesting. The quartic $(Q_e \widetilde Q_e)^2$ is always irrelevant, and thus drives the flow back to the original SCFT.
Finally, $Q_o \widetilde Q_o Q_e \widetilde Q_e$ is marginal, and moves the theory along the conformal manifold. 

\subsec{Higgsing}

We now consider giving a vev to a bifundamental $Q$,
\eqn\qvev{
\langle Q \rangle = v {\bf 1}_{N\times N}
}
If $Q$ is a bifundamental of $SU(N)_1 \times SU(N)_2$, this vev breaks $SU(N)_1 \times SU(N)_2 \rightarrow SU(N)_{diag}$. Being on the D-flat moduli space implies that $Q^i Q_j^\dagger - \widetilde Q^{\dagger i} \widetilde Q_j = c \delta^i_j$, so we do not need to give a vev to $\widetilde Q$ to solve the D-terms. However, if we choose to do so, giving a vev to $\widetilde Q$ just means giving a vev to a singlet of $SU(N)_{diag}$, which does not dramatically affect the resulting theory. 

When $SU(N) \times SU(N) \rightarrow SU(N)$, each bifundamental decomposes as $( {\bf N}, \overline {\bf N} ) \rightarrow ({\bf N}^2-{\bf 1}) + {\bf 1}$. One such adjoint gets eaten, but the remaining adjoint stays in the quiver. The resulting theory is a ${\cal S}_{\ell,k}$ theory with $\ell-1$ nodes and an adjoint chiral superfield attached to one of the nodes. In particular, if we give a vev to a $Q$ between nodes $k$ and $k+1$ in an ${\cal S}_\ell$ theory, the resulting theory is ${\cal S}_{\ell-1,k}$.

In Figure 14, we plot the difference of the central charges of ${\cal S}_{100}$ and ${\cal S}_{99,k}$ for several values of $k$.  The result is similar to what happens when we integrate out a bifundamental.  At large $N$, there are $a$-theorem violations.  This picture is generic when $\ell$ is even and independent of $k$.  

\bigskip
\centerline{{\epsfxsize=.5\hsize\epsfbox{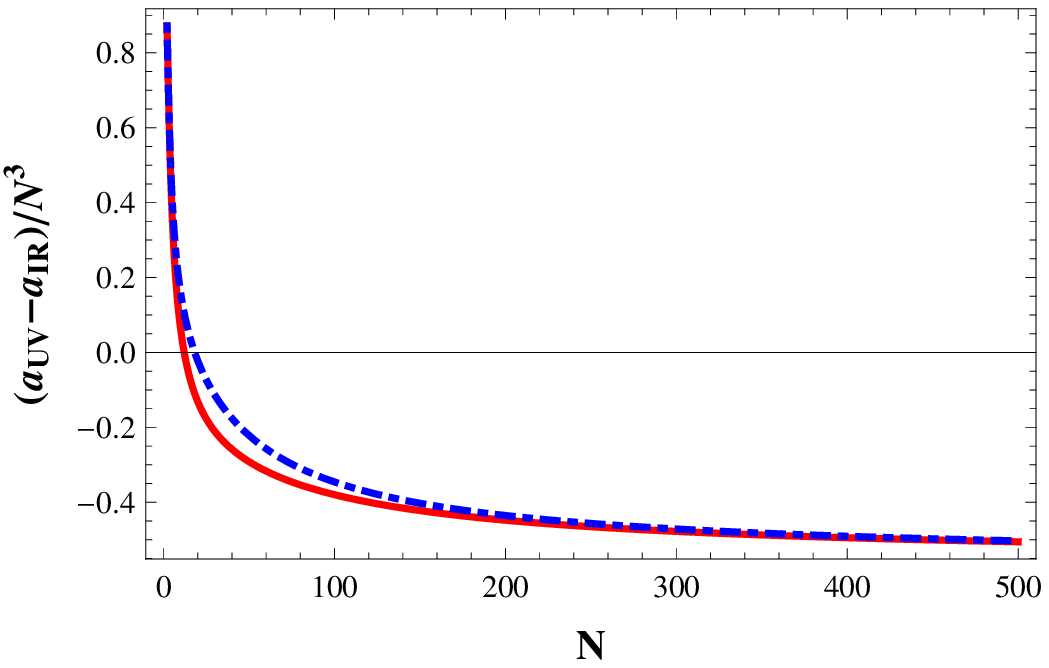}}}
\centerline{\ninepoint\sl \baselineskip=8pt {\bf Figure 14:}
{\sl Difference of central charges between ${\cal S}_{\ell}$ and ${\cal S}_{\ell-1,k}$ at $\ell=100$.}}
\centerline{\ninepoint\sl \baselineskip=8pt The colors correspond to $k=4$ (red, bottom) and 50 (dashed blue, top).} 
\bigskip

When $\ell$ is odd, the R-symmetry at large $N$ in the $T_N$ sector is the same for both ${\cal S}_{\ell}$ and ${\cal S}_{\ell-1,k}$.  In Figure 15, we plot the difference in central charges for $\ell =101$.  Similar to the case where we integrate out a bifundamental, in the large $N$ limit the difference in central charges will come from contributions of the hypermultiplets and vector multiplets.  These do not violate the $a$-theorem.  We do observe $a$-theorem violations at finite $N$, when $k \ll \ell$.

\bigskip
\centerline{{\epsfxsize=.5\hsize\epsfbox{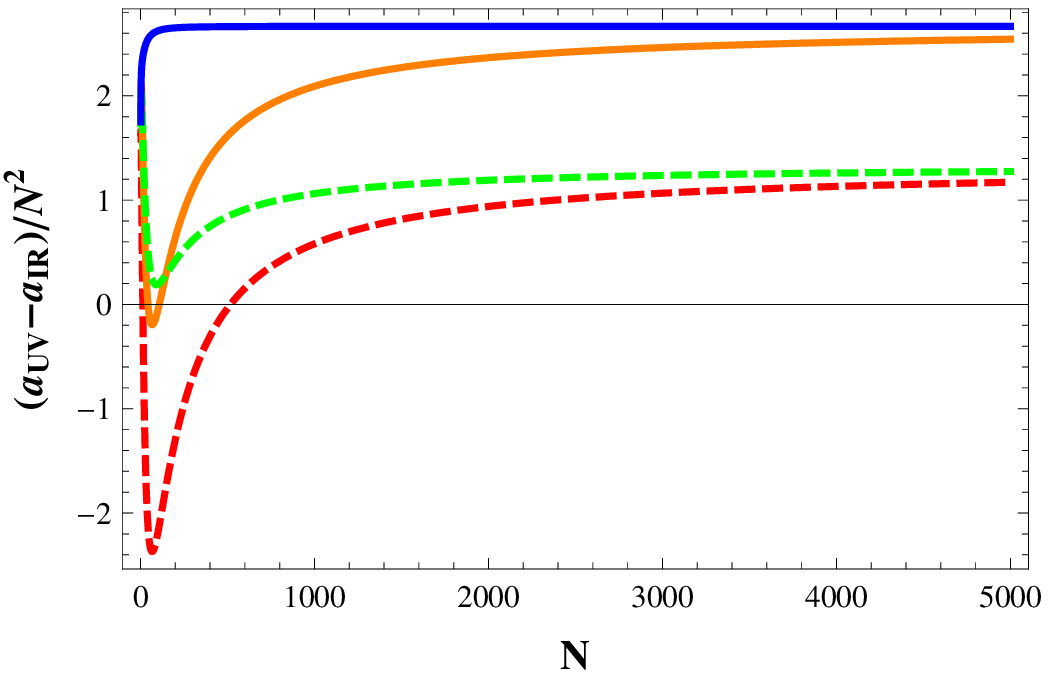}}}
\centerline{\ninepoint\sl \baselineskip=8pt {\bf Figure 15:}
{\sl Difference of central charges between ${\cal S}_{\ell}$ and ${\cal S}_{\ell-1,k}$ at $\ell=101$. }}
\centerline{\ninepoint\sl \baselineskip=8pt The color scheme is $k=3$ (dashed red, bottom), 10, 21, 50 (blue, top). }
\bigskip

Because of these $a$-theorem violations, it appears that the ${\cal S}_{\ell,k}$ cannot be the endpoint of flows from $\CS_\ell$ theories.  Since we cannot generate them dynamically, we also conclude that they do not exist. 

\subsec{Discussion}

In the previous subsections, we have seen many flows that violate the $a$-theorem, as well as some that do not. Since the $a$-theorem has now been proven, such flows cannot exist (and even beforehand, these flows would have seemed suspect). This leaves us with a puzzle: How are we to interpret the theories that show up at the endpoints, both UV and IR, of these flows?

Since both endpoints of the flow are strongly coupled, it is difficult to make definitive statements about whether or not these theories exist. This is generically true for strongly coupled SCFTs, and only in the presence of a duality (which we do not have here) can we hope for a stronger statement about the existence of a conformal phase. 

Still, we believe that the evidence presented above indicates that some of the theories we construct lead to good SCFTs while others do not. In particular, the ${\cal S}_\ell$ theories appear to be legitimate new SCFTs. For even $\ell$, the ${\cal S}_\ell$ theories are part of the conformal manifold of the $\CN=1$ Sicilian theories described in \BeniniMZ. The R-symmetry is $R_0$, and we do not change this by turning off the superpotential. Thus the even $\ell$ theories should be interacting SCFTs. From the UV point of view, it is reasonable (though by no means conclusive) to assume that the same is true for odd $\ell$ as well, even though the odd $\ell$ theories are no longer on the conformal manifold of the Sicilian theories of \BeniniMZ. However, the consistent flow from the theories in the present work to those of \BeniniMZ\ seems to indicate that these theories are well-behaved, and thus we posit that they exist as good SCFTs. A similar logic applies to the ${\cal S}_\ell^{\square}$ theories, since we can realize them by integrating out adjoints and subsequently going to the origin of the conformal manifold in the associated $\CN=2$ theory.

The theories with adjoints are more problematic. Since every dynamical realization we have tried in the present work, whether it be through Higgsing or via relevant deformations, leads to flows inconsistent with the $a$-theorem, we have several options. The first is that these theories simply do not have interacting conformal phases. Another is that there are some accidental symmetries that arise which we cannot see here. In either case, it is true that the theories as presented above are not SCFTs. In the absence of further evidence that they should exist, we therefore conclude that the theories with adjoints do not appear to be part of the landscape of $\CN=1$ SCFTs one can build from $T_N$ blocks. 

We thus arrive at the following categorization of the theories considered in this paper.
\eqn\theorylist{
\matrix{ {\cal S}_\ell & {\rm good} \cr
{\cal S}^{\square}_\ell & {\rm good}  \cr
{\cal S}^\circ_\ell & {\rm bad} \cr
{\cal S}_{\ell,k} & {\rm bad} }}

Since we find evidence that theories with adjoints and no superpotential all appear to be bad, it is worth asking why this is so. Naively, this is perhaps not surprising -- starting with an $\CN=2$ SCFT, one would not generically expect to be able to turn off the $Q X \widetilde Q$ terms without doing violence to the theory. For example, it may be that the gravity duals (i.e. the Maldacena-Nu\~nez solutions) of the $\CN=2$ theories become singular when we try to turn off the $\CN=2$ superpotential. It is far from clear that the theories we consider here should have gravity duals at all, and we plan to come back to this question in future work.

Another natural question to ask is what happens in the IR for the flows which look like they violate the $a$-theorem. We have provided evidence that the naive endpoints are not good theories, but if they are not the IR CFT, then where does the theory go? This is obviously a very interesting and important question for understanding the dynamics of the theories considered in this paper, and at present we do not have a good understanding of what the right IR phase is. One interesting possibility is that the theory arrives at a mixed phase (as in \refs{\CsakiUJ\GaiottoJF\CraigTX-\CraigWJ}). Another option is that, when we give mass to a bifundamental in an ${\cal S}_\ell$ theory, the entire theory becomes massive. We hope to return to this question in the future.

\newsec{Conclusions}

The main goal of this paper has been to take a small step in the exploration of new $\CN=1$ theories that one can build with $T_N$ blocks. It is not {\it a priori} obvious which such theories lead to interacting SCFTs and which do not. In the simplest case, where $T_N$'s are connected with hypermultiplets and intervening adjoints, our work indicates that only theories without adjoints are good SCFTs. However, this leaves open many questions and directions for future work.

The first question is whether one can extend the constructions presented here and build new SCFTs with $T_N$'s and hypermultiplets. For example, one potential avenue towards new SCFTs is to hook together the ${\cal S}_\ell$ theories. The exploration of such theories is presently underway \IbouTA, and the preliminary results suggest that the ${\cal S}_\ell$ theories can be connected in interesting ways, leading to a vast new landscape of $\CN=1$ SCFTs. 

Another interesting question is whether the theories presented here have gravity duals. If such dual theories exist, they should fall into the classification of \GauntlettZH, and show up via M5-branes wrapping an appropriate (punctured) Riemann surface in a $(2,0)$ six-dimensional theory. It would be interesting to determine, for example, the $\CN=1$ version of the story presented in \GaiottoGZ. 

\bigskip\bigskip

\centerline{\bf Acknowledgments}

We would like to thank Nikolay Bobev, Chris Beem, Tim Cohen, Henriette Elvang, Nick Hamalgyi, Ken Intriligator, Andrey Katz, Leo Pando-Zayas, David Simmons-Duffin, Phil Szepietowski, and Yuji Tachikawa for useful discussions and comments on the manuscript.  BW would additionally like to thank the Simons Center for Geometry and Physics, where part of this work was completed. IB is supported by DOE grant DE-FG02-95ER40899 and a University of Michigan Rackham Science Award. BW is supported by DOE grant DE-FG02-95ER40899 and the Fundamental Laws Initiative of the Center for the Fundamental Laws of Nature, Harvard University. BW is additionally supported by Eric Jankowski, emotionally.

 \listrefs
 
\end